\documentclass[journal=nalefd,manuscript=letter,layout=twocolumn,%
doi=true,email=false,maxauthors=0]{achemso}

\pdfpageattr {/Group << /S /Transparency /I true /CS /DeviceRGB>>}
\usepackage{graphicx}
\usepackage{amsmath,amssymb,bm}
\usepackage{hyperref}
\hypersetup{%
pdftitle={Exciton gas transport through nano-constrictions}, %
pdfauthor={C. Xu, J.R. Leonard, C.J. Dorow, L.V. Butov, M.M. Fogler, D.E. Nikonov, I.A. Young},%
pdfpagemode={UseNone},%
pdfstartview={FitH},%
breaklinks=true,%
citecolor=blue,%
colorlinks=true,%
linkcolor=blue,%
urlcolor=blue}

\usepackage[margin=0pt,font=scriptsize,labelfont=bf,%
labelsep=period,textfont=scriptsize]{caption}

\usepackage[T1]{fontenc}
\usepackage{mathpazo, mathabx}

\newcommand*{\myfontsize}{\footnotesize}
\newcommand*{\unit}[1]{\,\mathrm{#1}} 

\author{Chao Xu}
\email{chx035@ucsd.edu}
\author{J. R. Leonard}
\author{C. J. Dorow}
\author{L. V. Butov}
\author{M. M. Fogler}
\affiliation{Department of Physics, University of California San Diego, 9500 Gilman Drive, La Jolla, California 92093, USA}
\author{D. E. Nikonov}
\author{I. A. Young}
\affiliation{Components Research, Intel Corporation, Hillsboro, Oregon 97124, USA}

\date{\today}

\title{Exciton gas transport through nano-constrictions}

\abbreviations{QPC}


\begin{document}
\myfontsize
\begin{tocentry}
	
	%
	%
	%
	%
	\includegraphics[height=1.40in]{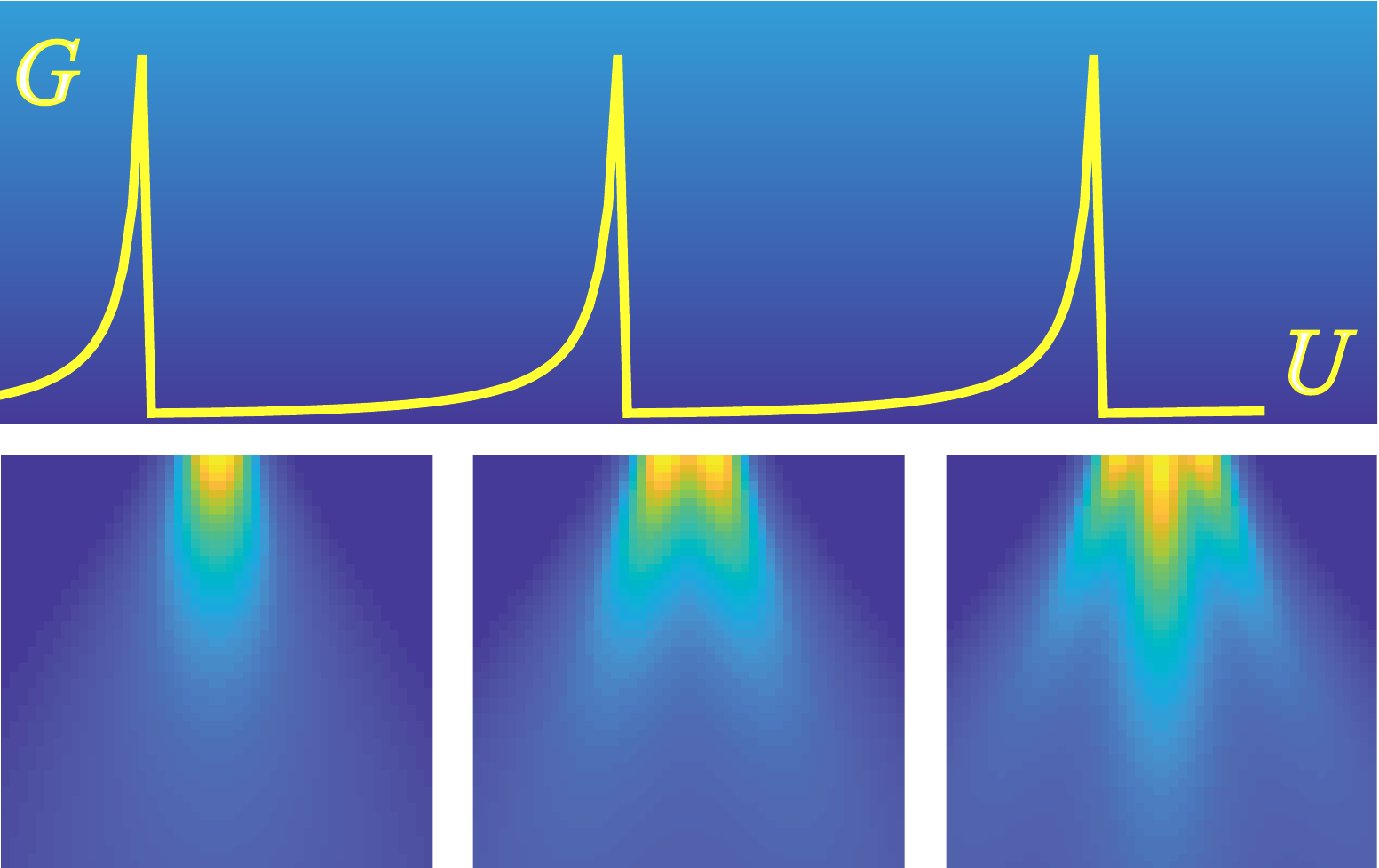}
	
	\scriptsize{}
	
\end{tocentry}

\begin{abstract}
	
	An indirect exciton is a bound state of  
	an electron and a hole in spatially separated layers.
	Two-dimensional indirect excitons can be created optically in heterostructures
	containing double quantum wells or atomically thin semiconductors.
	We study theoretically transmission of such bosonic quasiparticles through
	nano-constrictions.
	We show that quantum transport phenomena, e.g.,
	conductance quantization, single-slit diffraction, two-slit interference, and the Talbot effect
	are experimentally realizable in systems of indirect excitons.
	We discuss similarities and differences between
	these phenomena and their counterparts in electronic devices.
	
\end{abstract}

\myfontsize

\medskip

\begin{figure}
	\begin{center}
		\includegraphics[width=3.3in]{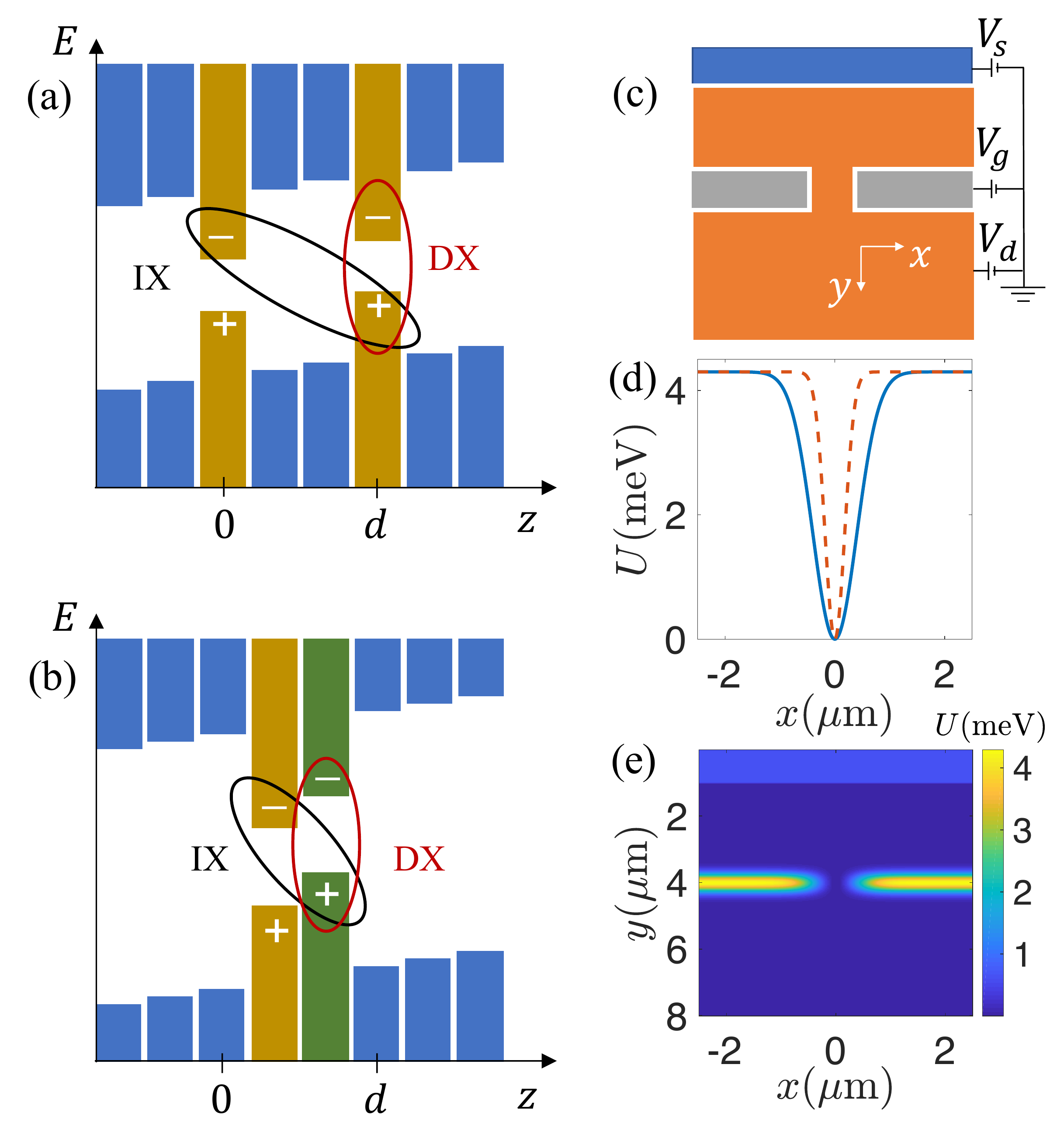}
		\caption{
			(a) Schematic energy band diagrams of an atomically thin electron-hole bilayer subject to an external electric field. The direct exciton (DX) is indicated by the red oval, the indirect exciton (IX) by the black one. The bars represent atomic planes.
			(b) Same as (a) for a spacer-free type II heterostructure.
			Various combinations of 2D materials can be used, examples include: for GaAs structures, GaAs (green), AlAs (yellow), AlGaAs (blue); for TMD type I structures, ${\rm{MoS_2}}$ (yellow), hBN (blue); for TMD type II structures, ${\rm{WSe_2}}$ (green), ${\rm{MoSe_2}}$ (yellow), hBN (blue)
			(c) Schematic of an IX QPC device.
			Voltage $V_s$ ($V_d$) controls the electrochemical potential of the source (drain);
			$V_g$ creates a potential constraining IXs from the sides.
			(d) Variation of the IX potential along the line $y = y_0$ across the QPC for
			$a_x = 180$ and $400\unit{nm}$ (dashed and solid trace, respectively) with
			other parameters as follows:
			$A = 4.30\unit{meV}$,
			$C = 1$,
			$a_y = 200\unit{nm}$,
			$U_{\mathrm{sd}} = 1.0\unit{meV}$,
			$y_0 = 4.0\,\mu\mathrm{m}$,
			$y_1 = 1.0\,\mu\mathrm{m}$,
			see Eq.~\eqref{eqn:U}.
			(e) False color map of $U(x, y)$ for the larger $a_x$ in (d) .
		}
		\label{fig:setup1}
	\end{center}
\end{figure}

\textbf{Introduction}.
Indirect excitons (IXs) in coupled quantum wells have emerged as a new platform for investigating quantum transport.
The IXs have long lifetime,~\cite{lozovik1976new} long propagation distance,~\cite{hagn1995electric, gartner2006drift, hammack2009kinetics, leonard2009spin, 
	lazic2010exciton, alloing2012nonlinear, lazic2014scalable, finkelstein2017transition}
and long coherence length
at temperatures $T$ below the temperature $T_0$ of quantum degeneracy [Eq.~\eqref{eqn:mu}].{\color{blue}~\cite{high2012spontaneous}} 
Although an IX is overall charge neutral,
it can couple to an electric field via its static dipole moment $e d$,
where $d$ is the distance between the electron and hole layers [Fig.~\ref{fig:setup1}(a)].
These properties enable experimentalists to 
study transport of quasi-equilibrium IX systems subject to artificial potentials
controlled by external electrodes [Fig.~\ref{fig:setup1}(b)].

\begin{figure}
	\begin{center}
		\includegraphics[width=3.3in]{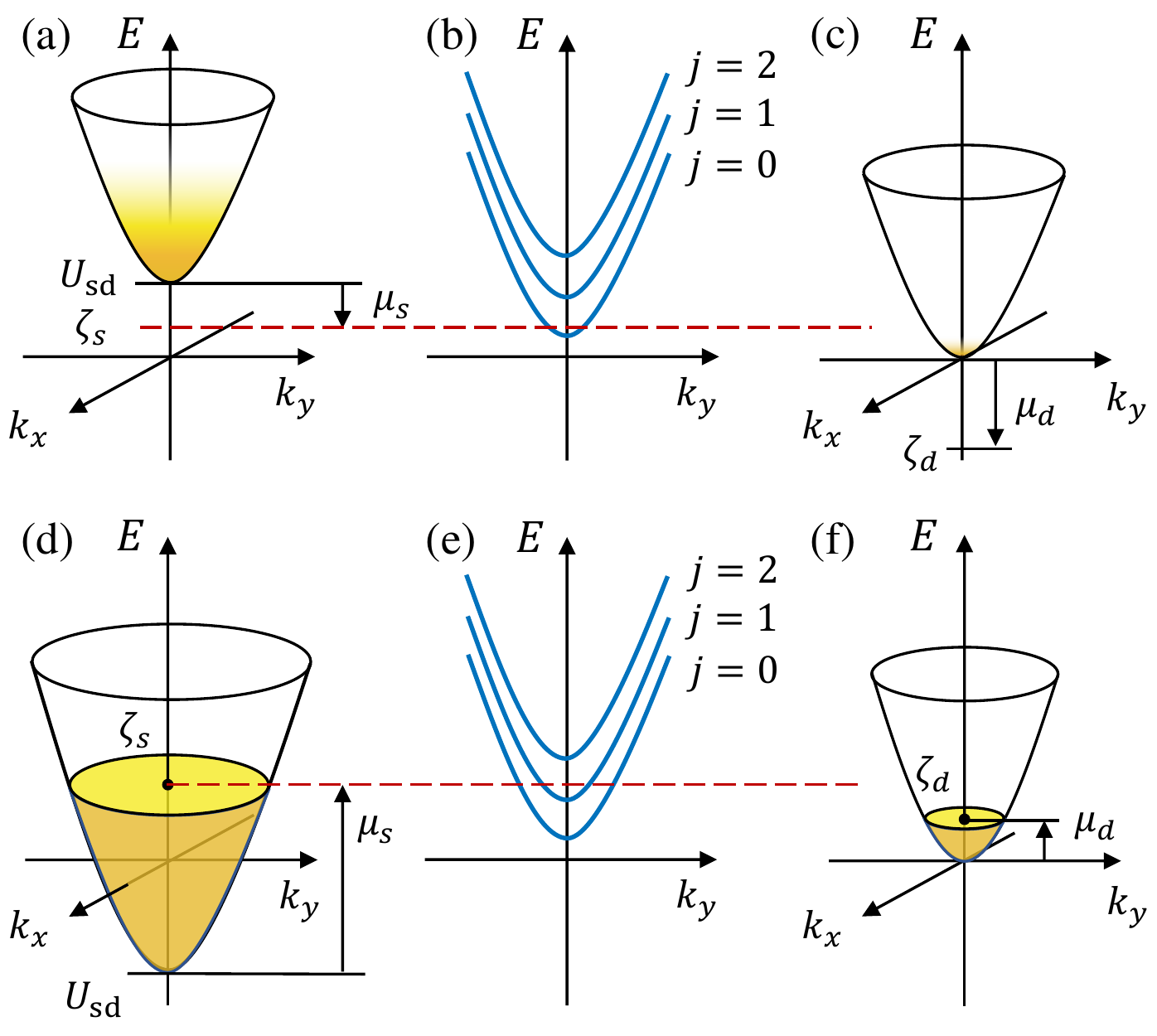}
		\caption{
			Energy spectra at the source (a, d), the QPC (b, e), and the drain (c, f).
			The electrochemical potential of the source is indicated by
			the horizontal dashed line.
			The shading represents the occupation factor.
			Panels (a-c) are for bosonic particles, such as IXs.
			Panels (d-f) are for fermions, e.g., electrons.
		}
		\label{fig:energy_diagrams}
	\end{center}
\end{figure}

In this Letter, we examine theoretically
the transport of IXs through nano-constrictions [Fig.~\ref{fig:setup1}(c-e)].~\cite{dorow2018split}
Historically, studies of transmission of particles 
through narrow constrictions have led
to many important discoveries.
In particular, investigations of electron transport through so called quantum-point contacts (QPCs)
have revealed that at low $T$ the conductance of smooth, electrostatically defined QPCs
exhibits step-like behavior as a function of their width.~\cite{wharam1988one, van1988quantized}
The steps appear in integer multiples of $ N e^2 / h$
(except for the anomalous first one~\cite{Micolich2013}) because
the electron spectrum in 
the constriction region is quantized into one-dimensional (1D) subbands
[Fig.~\ref{fig:energy_diagrams}(e)].
Here $N$ is the spin-valley degeneracy.
Associated with such subbands are current density profiles
analogous to the diffraction patterns of light passing through a narrow slit.
These patterns have been observed by nanoimaging of the electron flow.~\cite{topinka2000imaging}
The conductance quantization has also been observed in another tunable
fermionic system, a cold gas of $^6\mathrm{Li}$ atoms.~\cite{krinner2015observation}

For the transport of bosonic IXs, we have in mind
a conventional in solid-state physics setup where
the QPC is connected to the source and drain reservoirs of 
unequal electrochemical potentials, $\zeta_s$ and $\zeta_d$
[Fig.~\ref{fig:energy_diagrams}(a-c)].
The difference $\zeta_s - \zeta_d > 0$ is analogous to the source-drain voltage in electronic devices.
Whereas electrons are fermions, IXs behave as bosons.
This makes our transport problem unlike the electronic one.
The problem is also different from the slit diffraction of photons
or other bosons, such as phonons, considered so far. 
Indeed, one cannot apply a source-drain voltage to photons or phonons in any usual sense.
[However, quantized \emph{heat} transport of phonons~\cite{schwab2000measurement, gotsmann2013quantized, cui2017quantized} has been studied.]

To highlight the qualitative features, we do our numerical calculations for the case where the drain side is empty, $\zeta_d = -\infty$.
The conductance of the QPC can be described by the bosonic variant of the standard Landauer-B\"uttiker theory.~\cite{Buttiker1985}
It predicts that the contribution of a given subband to the total conductance can exceed $N / h$ if
its Bose-Einstein occupation factor is larger than unity.~\cite{papoular2016quantized}
To the best of our knowledge,
there have been no direct experimental probes of this prediction in bosonic systems.
The closest related experiment is probably the study of $^6\mathrm{Li}$ atoms passing through a QPC in the regime of enhanced attractive interaction.~\cite{Krinner2016}
That experiment has demonstrated the conductance exceeding $N / h$ and 
the theory~\cite{Kanasz2016,Shun2017,Boyang2017} has attributed this excess to the virtual pairing of fermionic atoms into bosonic
molecules by quantum fluctuations. 

Below we present our theoretical results for the IX transport through single, double, and multiple QPCs.
We ignore exciton-exciton interaction
but comment on possible interaction effects at the end of the Letter.

\textbf{Model of a QPC}.
In the absence of external fields, the IXs are free to move in a
two-dimensional (2D) $x$--$y$ plane.
When an external electric field $E_z = E_z(x, y)$ is applied
in the $z$-direction [Fig.~\ref{fig:setup1}(a)],
an IX experiences the energy shift $U = -e E_z d$.
This property makes it possible to create desired external potentials $U(x, y)$ acting on the IXs.
To engineer a QPC, $U(x, y)$ needs to have a saddle-point shape.
Such a potential can be created using a configuration of electrodes:
a global bottom gate plus a few local gates on top of the device.
As depicted in Fig.~\ref{fig:setup1}(b),
two of such top electrodes (gray) can provide the lateral confinement and another two
(blue and orange) can control the potential at the source and the drain.
More electrodes can be added if needed.
Following previous work,~\cite{dorow2018split}
in our numerical simulations we use a simple model for $U(x, y)$:
\begin{equation}
\begin{split}
U(x, y) &= A
\left(1 - C e^{-{x^2} / {2 a_x^2}}\right)
e^{-{(y - y_0)^2} / {2 a_y^2}}
\\
&+ U_{\mathrm{sd}} f_F\left(\frac{y - y_1}{s}\right),
\label{eqn:U}
\end{split}
\end{equation}
where $f_F(x) = (e^x + 1)^{-1}$ is the Fermi function.
The width parameters $a_x$ and $a_y$ and the coefficients $A$ and $C$ on the first line of Eq.~\eqref{eqn:U} are tunable by the gate voltage $V_g$ [Fig.~\ref{fig:setup1}(b)].
The second line in Eq.~\eqref{eqn:U} represents a gradual potential drop
of magnitude $U_{\mathrm{sd}}$, central coordinate $y_1$,
and a characteristic width $s$ in the $y$-direction.
These parameters are controlled by voltages $V_s$ and $V_d$.
Examples of $U(x, y)$ are shown in Fig.~\ref{fig:setup1}(c,d).

For qualitative discussions, we also consider the model of a quasi-1D channel of a long length $L_c$
and a parabolic confining potential.
The corresponding $U(x, y)$ is obtained by replacing the top line
of Eq.~\eqref{eqn:U} with
\begin{equation}
f_F\left(\frac{y_0 - L_c - y}{s}\right) f_F\left(\frac{y - y_0}{s}\right)
\left(U_0 + \frac{1}{2} U_{xx} x^2\right),
\label{eqn:U_quadratic}
\end{equation}
where $U_{xx} = A C / a_x^2$.
The energy subbands [Fig.~\ref{fig:energy_diagrams}(b)] in such a channel are
\begin{equation}
E_j(k_y) = U_0 + \hbar \omega_x \left(j + \frac12\right) + \frac{\hbar^2 k_y^2}{2 m}\,,
\quad
j = 0, 1, \ldots
\label{eqn:E_j}
\end{equation}
The associated eigenstates are the products of the plane waves $e^{i k_y y}$
and the harmonic oscillator wavefunctions
\begin{equation}
\chi_j(x) = H_j \left({\sqrt2\,x}/{w_0}\right)\exp \left(-{x^2} / {w_0^2}\right),
\label{eqn:chi_j}
\end{equation}
where $H_j(x)$ is the Hermite polynomial of degree $j$.
The length $w_0$ and frequency $\omega_x$ are given by
\begin{equation}
w_0 = \sqrt{\frac{2\hbar}{m \omega_x}}\,,
\quad
\omega_x = \sqrt{\frac{U_{xx}}{m}} \,.
\label{eqn:w_0}
\end{equation}
The subband energy spacing $\hbar \omega_x$ can be estimated as
\begin{equation}
\hbar \omega_x \simeq 8.8\unit{meV} 
\left(\frac{m_0}{m}\right)^{1/2} \left(\frac{A C}{\mathrm{meV}}\right)^{1/2}
\left(\frac{\mathrm{nm}}{a_x}\right),
\end{equation}
which is approximately $0.1\unit{meV}$ for
devices made of transition metal dichalcogenides (TMDs),
where exciton effective mass $m = 1.0 m_0$ and $a_x = 180\unit{nm}$.
Hence, the onset of quantization occurs at $T \sim 1\unit{K}$.
The same characteristic energy scales
in GaAs quantum well devices,~\cite{dorow2018split} can be obtained in a wider
QPC, $a_x = 400\unit{nm}$, taking advantage of the lighter mass, $m = 0.2 m_0$,
cf.~solid and dashed lines in Fig.~\ref{fig:setup1}(d).
In both examples $\hbar \omega_x$ is much smaller than the IX binding energy
($\sim 300\unit{meV}$ in TMDs~\cite{Latini2017iea} and $\sim 4\unit{meV}$ in GaAs~\cite{Szymanska2003ebi, Sivalertporn2012dai}).
Therefore, we consider the approximation ignoring internal dynamics of the IXs as they pass through the QPC~\cite{Grasselli2016etb} and treat them as point-like particles.
Incidentally, we do not expect any significant reduction of the exciton binding energy due to many-body screening in the considered
low-carrier-density regime in semiconductors with a sizable gap and no extrinsic doping.
(At high carrier densities, the screening effect can be substantial.\cite{Kharitonov2008})

\noindent\textbf{Conductance of a QPC}.
As with usual electronic devices, we imagine that our QPC is connected to semi-infinite source and drain leads (labeled by $l = s, d$).
Inside the leads the IX potential energy $U(x, y)$
tends to asymptotic values $U_l$.
Without loss of generality, we can take $U_d = 0$, as in Eq.~\eqref{eqn:U}.
The difference $U_{\mathrm{sd}} = U_s - U_d = \alpha (e V_s - e V_d)$ is a linear function of 
the control voltages $V_s$, $V_d$ [Fig.~\ref{fig:setup1}(b)].
The coefficient of proportionality can be estimated as $\alpha \sim d / d_g$,
where $d_g$ is the vertical distance between the top and bottom electrodes.
The IX energy dispersions at the source and the drain are parabolic
and are shifted by $U_{\mathrm{sd}}$ with respect to one another,
see Fig.~\ref{fig:energy_diagrams}(a, c).
In the experiment, the IX density of the reservoirs can be controlled by photoexcitation power.
For example, the IX density $n_s$ can be generated on the source side only,~\cite{dorow2018split}
while the drain side can be left practically empty, $n_d \ll n_s$.
This is the case we focus on below.
The chemical potentials $\mu_l$ are related to the densities $n_l$ via~\cite{Remeika2015,Ivanov1999}
\begin{equation}
\mu_l = T \ln\left(1 - e^{-T_0 / T}\right)\,,
\quad T_0 \equiv \frac{2\pi \hbar^2 n_l}{m N}\,.
\label{eqn:mu}
\end{equation}
Note that we count the chemical potentials from the minima of the appropriate energy spectra and that we use the units system $k_B \equiv 1$.
In turn, the electrochemical potentials are given by
\begin{equation}
\zeta_l = \mu_l + U_l\,,
\label{eqn:zeta}
\end{equation}
which implies
\begin{equation}
U_{\mathrm{sd}} = (\zeta_s - \zeta_d) - (\mu_s - \mu_d)\,.
\label{eqn:U_sd}
\end{equation}
In the terminology of electron devices, the first term in Eq.~\eqref{eqn:U_sd} is
related to the source-drain voltage $V_{\mathrm{sd}}$, \latin{viz}.,
$\zeta_s - \zeta_d \equiv -e V_{\mathrm{sd}}$.
The second term in Eq.~\eqref{eqn:U_sd} is referred to as the built-in potential,
which is said to originate from charge redistribution in the leads.

If the particle densities $n_l$ are fixed,
variations of $\zeta_s - \zeta_d$ rigidly track those of $U_{\mathrm{sd}}$.
The differential conductance can be computed
by taking the derivative of the source-drain particle current $I$ with respect to 
$U_{\mathrm{sd}}$:
\begin{equation}
G = \left({\partial I} / {\partial U_{\mathrm{sd}}}\right)_{n_s, n_d}.
\end{equation}
$G$ has dimension of $1 / h$ and its natural quantum unit is $N / h$.
(The spin-valley degeneracy is $N = 4$ in both TMDs and GaAs.)

\begin{figure}
	\begin{center}
		\includegraphics[width=3.3in]{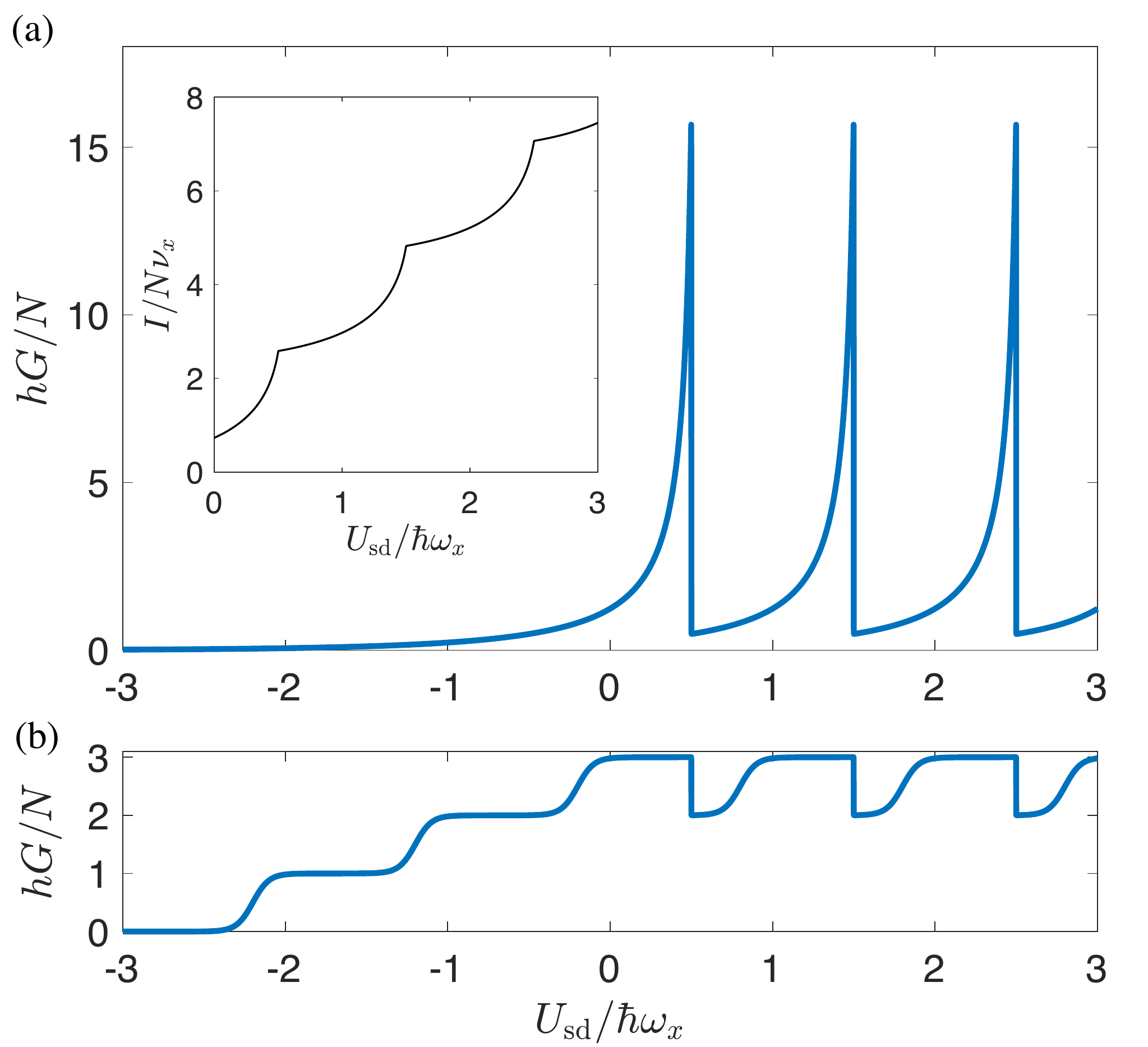}
		\caption{
			(a) Conductance of a bosonic QPC \latin{vs}. $U_{\mathrm{sd}}$
			from the adiabatic approximation [Eq.~\eqref{eqn:G}]
			for $T = 0.8 \hbar\omega_x$, $\mu_s = -0.05 \hbar\omega_x$.
			Inset: total current as a function of $U_{\mathrm{sd}}$.
			The notation used on the $y$-axis is $\nu_x \equiv \omega_x /\, 2\pi$.
			(b) Conductance of a fermionic QPC for
			$T = 0.05 \hbar\omega_x$, $\mu_s = 2.7 \hbar\omega_x$.
			Note that the number of conductance steps occurring at $U_{\mathrm{sd}} < E_0^0 = \frac12 \hbar\omega_x$ is equal to $\mu_s / \hbar\omega_x$ rounded up to the nearest integer.
		}
		\label{fig:G}
	\end{center}
\end{figure}

For the long-channel model [Eq.~\eqref{eqn:U_quadratic}],
$G$ can be calculated analytically because
the single-particle transmission coefficients
through the QPC have the form $t_j(E) = \theta\bigl(E - E_j^0\bigr)$, i.e., they take values of either $0$ or $1$.
Here $E_j^0 \equiv E_j(0)$ and $\theta(x)$ is the Heaviside step-function.
Adopting the standard Landauer-B\"uttiker theory for fermions~\cite{Buttiker1985} to the present case of bosons (see also~\cite{papoular2016quantized}),
we find the total current to be $I = \sum_j (I_{s,j} - I_{d,j})$
where the partial currents are
\begin{equation}
I_{a,j} = \frac{N}{h} \int\limits_{U_a}^\infty f_B\biggl(\frac{E - \zeta_a}{T}\biggr)
t_j(E) d E
\end{equation}
and $f_B(x) = (e^x - 1)^{-1}$ is the Bose function.
In turn, the conductance is
\begin{equation}
G = \frac{N}{h} \sum_j \theta\bigl(E_j^0 - U_{\mathrm{sd}}\bigr)
f_B\biggl(\frac{E_j^0 - U_{\mathrm{sd}} - \mu_s}{T}\biggr).
\label{eqn:G}
\end{equation}
The dependence of $G$ on $U_{\mathrm{sd}}$ is illustrated in Fig.~\ref{fig:G}(a). 
The conductance exhibits asymmetric peaks.
Each peak signals the activation of a new conduction channel whenever $U_{\mathrm{sd}}$ approaches the
bottom of a particular 1D subband.
All the peaks have the same shape, which is the mirror-reflected Bose function with a sharp cutoff.
As $T$ decreases, the width of the peaks $\Delta \sim \min (T, -\mu_s)$ decreases. 
The magnitude of the peaks $G_{\mathrm{max}}
= ({N} / {h}) f_B\left(-{\mu_s} / {T}\right)$ can be presented in the form
\begin{equation}
G_{\mathrm{max}} = \frac{N}{h} f_{\mathrm{max}}\,,
\label{eqn:G_max}
\end{equation}
where $f_{\mathrm{max}}$ is the occupation of the lowest energy state at the source.
For fermions, $f_{\mathrm{max}}$ is limited by $1$.
For 2D bosons, $f_{\mathrm{max}} = e^{T_0 / T} - 1$ exceeds $1$ and, in turn, $G_{\mathrm{max}}$ exceeds $N / h$ at $T < T_0 /\! \ln 2$.

\begin{figure*}
	\centering
	\includegraphics[width=5.50in]{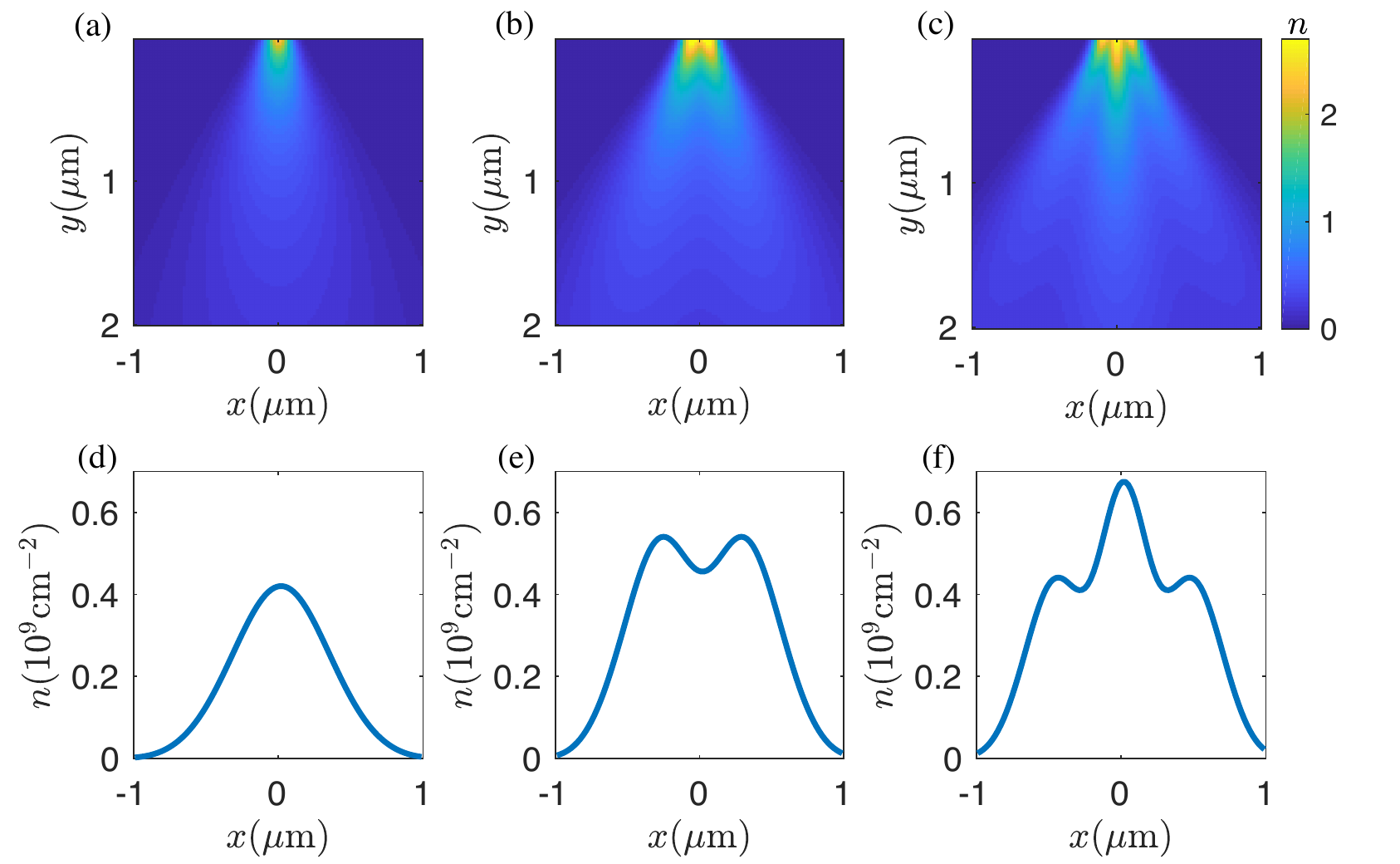}
	\caption{
		The IX density distribution in a QPC for three different $U_{\mathrm{sd}}$. 
		Panels (a-c) depict the density distribution $n$ in the $x$--$y$ plane in units of $10^9\unit{cm}^{-2}$. 
		Panels (d-f) show the profiles of $n$ along the midline, $y = 1\,\mu\mathrm{m}$, of the region plotted directly above.
		Values of $U_{\mathrm{sd}}$ are adjusted to highlight the contributions of particular subbands:
		(a, d) $j = 0$ for $U_{\mathrm{sd}} = 0.52$,
		(b, e) $j = 1$ for $U_{\mathrm{sd}} = 1.52$,
		(c, f) $j = 2$ for $U_{\mathrm{sd}} = 2.52$,
		all in units of $\hbar \omega_x = 0.10\unit{meV}$; $T = 1.0\unit{K}$ everywhere.
	}
	\label{fig:n2}
\end{figure*}

The sudden drops of $G$ at $U_{\mathrm{sd}} = E_j^0$ [Fig.~\ref{fig:G}(a)]
occur because the current carried by each subband saturates as
soon as it becomes accessible to all the IXs
injected from the source, down to the lowest energy $U_{\mathrm{sd}}$.
These constant terms do not affect the differential conductance $G$.
The total current as a function of $U_{\mathrm{sd}}$ is plotted in the inset of
of Fig.~\ref{fig:G}(a).

It is again instructive to compare these results with the more familiar ones for fermions,
which are obtained replacing the Bose function $f_B$ with the Fermi function $f_F$ in
Eq.~\eqref{eqn:G}.
In the regime where $U_{\mathrm{sd}}$ is large and negative,
the Heaviside functions in Eq.~\eqref{eqn:G} play no role,
so that $G$ traces the expected quantized staircase shown in Fig.~\ref{fig:G}(b).
The conductance steps occur whenever $U_{\mathrm{sd}} = E_j^0 - \mu_s$.
Once $U_{\mathrm{sd}}$ approaches $E_0^0 = \hbar\omega_x / 2$, a different behavior is found:
the conductance displays additional sudden drops at $U_{\mathrm{sd}} = E_j^0$,
which causes it to oscillate between two quantized values.
These drops appear for the same reason as in the bosonic case:
current saturation for each subband that satisfies the condition $E_j^0 < U_{\mathrm{sd}}$.

The adiabatic QPC model considered above is often a good approximation~\cite{van1988quantized, wharam1988one}
for more realistic models, such as Eq.~\eqref{eqn:U}.
The latter cannot be treated analytically but we were able
to compute $G$ numerically,
using the transfer matrix method~\cite{Usuki1995nao} (see below).
The results are presented in the Supporting Information.
The main difference from the adiabatic case is that
the conductance peaks are reduced in magnitude and broadened.

\textbf{Density distribution in a QPC}.
In analogy to experiments with electronic QPCs,~\cite{topinka2000imaging}
it would be interesting to study quantized conduction channels by optical imaging of IX flow.
Unlike electrons, IXs can recombine and emit light, so that the IX photoluminescence can be used for measuring their current paths.
This motivates us to model the density distribution $n(x, y)$ of IXs in the QPC.
We begin with analytical considerations and then present our numerical results,
Fig.~\ref{fig:n2}.

The calculation of the $n(x, y)$ involves two steps.
First, we solve the Schr\"odinger equation
for the single-particle energies $E = \hbar^2 k^{\prime 2} / (2m)$ and the wavefunctions $\psi(x, y)$ of IXs subject to the potential $U(x, y)$.
The boundary conditions for the $\psi(x, y)$ is
to approach linear combinations of plane waves
of momenta $\mathbf{k}^{\prime} = (k_x^{\prime}, \pm k_y^{\prime})$ at $y \to -\infty$ but contain no
waves of momenta $\mathbf{k}^{\prime \prime} = (k_{x}^{\prime \prime}, k_{y}^{\prime \prime})$ with $\mathrm{Re}\,k_{y}^{\prime \prime} < 0$ at $y \to +\infty$.
Such states correspond to waves incident from the source.
Second, we sum the products
$f_B(E - \zeta_s) |\psi(x, y)|^2$ to obtain the $n(x, y)$.
We choose not to multiply the result by $N$,
so that our $n(x, y)$ is the IX density per spin per valley.

For a qualitative discussion, let us concentrate on the simplest case of large $k_y^{\prime}$, i.e., fast particles.
As the incident wave propagates from the source lead, it first meets the smooth potential drop $U_{\mathrm{sd}}$ at $y = y_1$.
Assuming $k_y^{\prime} s \gg 1$,
the over-the-barrier reflection at $y = y_1$ can be neglected.
Therefore, the $y$-momentum increases to the value $k_y$
dictated by the energy conservation,
\begin{equation}
k_y^{2} = k_y^{\prime 2} + ({2 m} / {\hbar^2}) U_{\mathrm{sd}}\,,
\label{eqn:k_y_prime}
\end{equation}
while the $x$-momentum $k_x^{\prime}$ remains the same.
Subsequently, the incident wave impinges on the QPC at $y = y_0$.
Typically, this causes a strong reflection back to the source.
However, certain eigenstates have a nonnegligible transmission.
After passing the QPC, their wavefunctions
expand laterally with the characteristic angular divergence of $1 / (k_y w_0) \ll 1$.
Such wavefunctions can be factorized
$\psi(x, y) = e^{i k_y y} \chi(x, y)$, where the slowly varying amplitude $\chi(x, y)$
obeys the eikonal (or paraxial) equation
\begin{equation}
\left(-i \hbar v_y {\partial_y} - \frac{\hbar^2}{2 m}\, {\partial_x^2} + U\right)\chi = 0\,,
\quad
v_y \equiv \frac{\hbar k_y}{m}\,.
\label{eqn:eikonal}
\end{equation}
If the model of the long constriction [Eq.~\eqref{eqn:U_quadratic}] is a good approximation,
the solution is as follows.
Inside the QPC, $\chi({\mathbf{r}})$ is proportional to a particular
oscillator wavefunction $\chi_j(x)$ [Eq.~\eqref{eqn:chi_j}].
Outside the QPC, it behaves as a Hermite-Gaussian beam
whose probability density can be written in the scaling form
\begin{align}
|\chi(x, y)|^2 &= \frac{1}{b(y)}\, \chi_j^2 \left(\frac{x}{b(y)}\right),
\label{eq:chi}\\
b(y) &= \sqrt{1 + \frac{(y - y_0)^2}{w_y^2}}\,,
\label{eq:b}\\
w_y &= \frac12\, k_y w_0^2\,.
\label{eq:w_y}
\end{align}
This representation has been used in the study of a Bose-Einstein condensate (BEC) expansion from a harmonic trap.~\cite{kagan1996evolution, castin1996bose}
Our problem of the steady-state 2D transport through the QPC maps to the noninteracting limit of this problem in $1+1$D,
with $t = (y - y_0) / v_y$ playing the role of time.
In other words, the IX current emerging from the QPC is mathematically similar to a freely expanding BEC.
This mapping is accurate if the characteristic width of the energy distribution of the IX injected into the QPC from the source is small enough,
$\Delta \ll \hbar \omega$.
Dimensionless function $b(y)$ has the meaning of the expansion factor.
A salient feature of the eigenfunctions are the nodal lines $\chi(x, y) = 0$.
The lowest subband $j = 0$ has no such lines whereas the higher subbands have exactly $j = 1, 2, \ldots $ of them. 

For a quantitative modeling, we carried out simulations using the transfer matrix method.~\cite{Usuki1995nao}
This method gives a numerical solution
of the Schr\"odinger equation discretized on a finite-size real-space grid
for a given energy $E$
and the boundary conditions described above.
To obtain the total particle density, we summed
the contributions of individual states, making sure to include enough $E$'s to achieve convergence.
In all our calculations the chemical potential $\mu_s$ was fixed to
produce the density (per spin per valley) $n_s = 1.0\times 
10^{10}\unit{cm}^{-2}$ at the source.
Examples of such calculations
for temperature $T = 1.0\unit{K}$ are shown in Fig.~\ref{fig:n2}.
Since the partial densities $|\psi(x, y)|^2$ are weighted with the Bose-Einstein factor $f_B(E - \zeta_s)$,
the lowest-energy $j = 0$ subband typically dominates the total density,
making it look like a nodeless Gaussian beam [Fig.~\ref{fig:n2}(a,d)].
However, if $U_{\mathrm{sd}}$ is tuned slightly above the bottom of the $j = 1$ subband,
the contribution of this subband is greatly enhanced by the van Hove singularity of the 1D density of states $N / (\pi \hbar v_y) \propto \bigr(E - E_j^0\bigl)^{-1 / 2}$ inside the QPC.
As a result, $n(x, y)$ develops a valley line (local minimum) at $x = 0$,
which is the nodal line of function $\chi_1(x / b, y)$ [Fig.~\ref{fig:n2}(b,e)].
A similar phenomenon occurs when we tune $U_{\mathrm{sd}}$ to slightly above the bottom of $j = 2$ subband. Here
the density exhibits two valley lines, which approximately follow the nodal lines of function $\chi_2(x / b, y)$ [Fig.~\ref{fig:n2}(c,f)].
These theoretical predictions may be tested by imaging IX emission with
high enough optical resolution.
Note that these van Hove singularities do not enhance the differential conductivity $G$
because of the cancellation between the density of states and the particle velocity $v_y$.
As explained above, this leads to current saturation and thus negligible contribution of $j$th subband to $G$ at $U_{\mathrm{sd}} > E_j^0$.

\begin{figure}
	\begin{center}
		\includegraphics[width=2.30in]{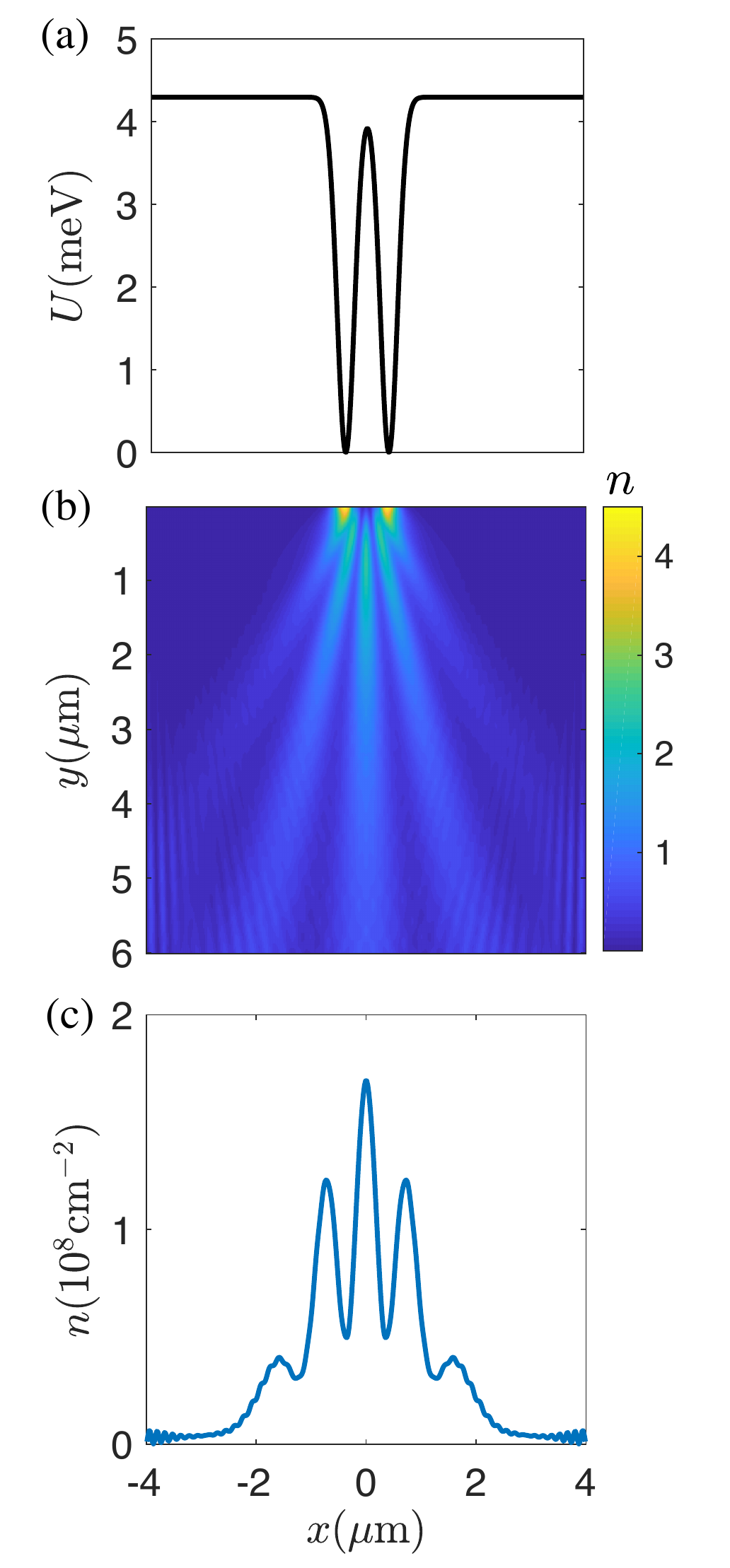}
		\caption{
			Double-QPC IX device.
			(a) The profile of the IX potential across the two constrictions. 
			(b) The IX density in units of $10^8\,\mathrm{cm}^{-2}$.
			(c) The IX density along the line $y = 2\,\mu\mathrm{m}$ in (b).
			The ripples at the flanks of the fringes are finite-size artifacts. 
		}
		\label{fig:2slits}
	\end{center}
\end{figure}

\textbf{Double and multiple QPC devices}.
We have next explored the exciton analog of the Young double-slit interference.
A short distance downstream from the first QPC
($1.0\,\mu\mathrm{m}$ along $y$), we
added another potential barrier constructed
from two copies of the single-QPC potential [Eq.~\eqref{eqn:U}]
shifted laterally in $x$.
As in Young's classical setup,
the first QPC 
plays the role of a coherent source for the double-QPC.
An example of the latter
with the center-to-center separation of $800\unit{nm}$
is shown in Fig.~\ref{fig:2slits}(a).
We computed the IX density distribution in this system for
$U_{\mathrm{sd}}$ at which the IXs fluxes
through all the QPCs are dominated by the $j = 0$ subband.
The IXs transmitted through the double-QPC
create distinct interference fringes, Fig.~\ref{fig:2slits}(b,c). 

\begin{figure}
	\begin{center}
		\includegraphics[width=2.80in]{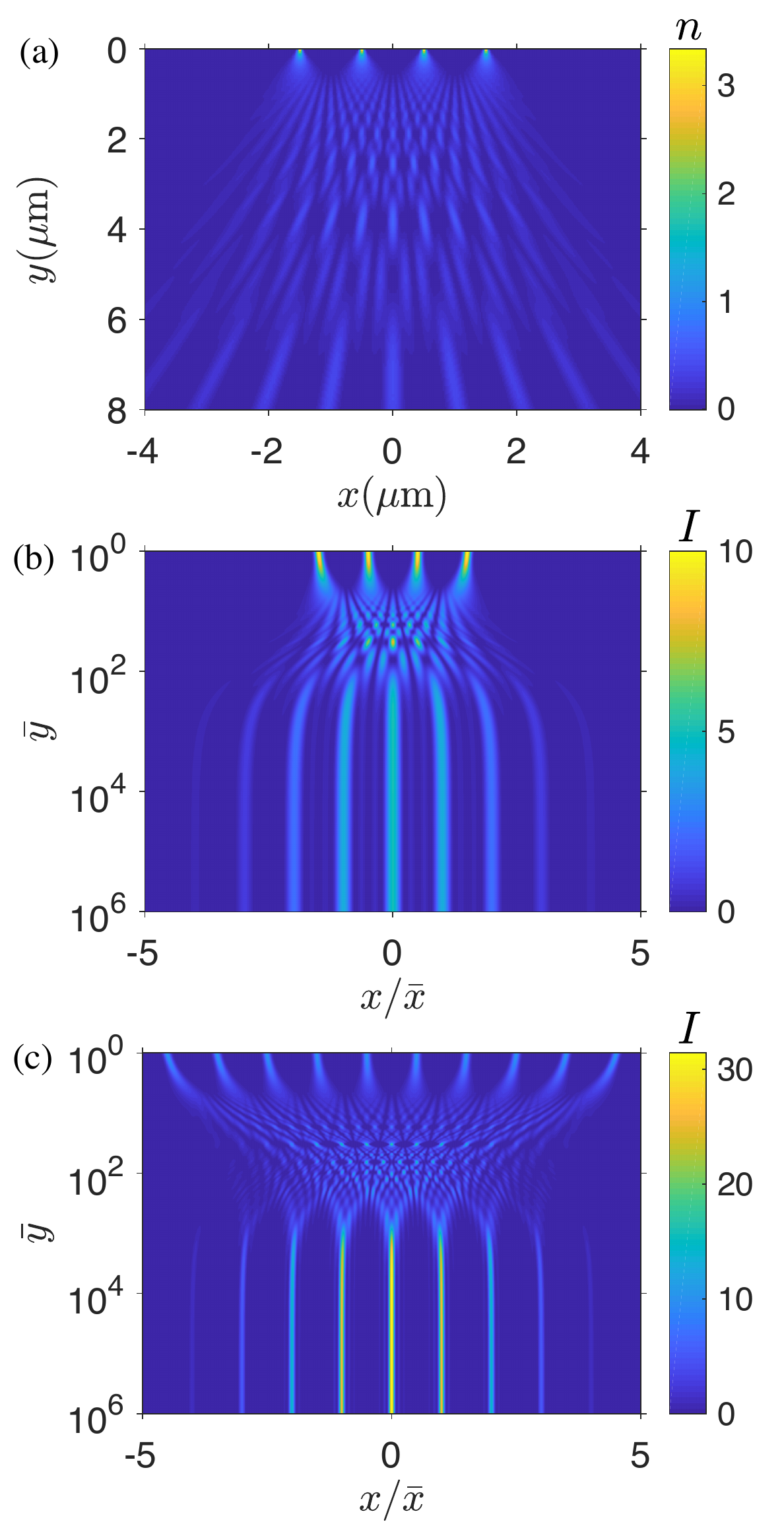}
		\caption{
			False color maps illustrating IX density in multiple QPC systems.
			(a) Talbot-like interference in a four-QPC device. 
			The incident beam consists of a single plane wave, corresponding to
			the $T = 0$ limit. $U_{\mathrm{sd}} = 0.10\unit{meV}$.
			(b) Rescaled intensity $I \equiv |\psi(\mathrm{r})|^2 \bar{x}$ of the interference pattern of $N_s = 4$ Gaussian beams.
			Following an example in the literature,~\cite{Juffmann2013} the horizontal axis is
			normalized by $\bar{x} = (\lambda y / a) + a$ with $a = 1\,\mu\mathrm{m}$.
			The vertical axis represents $\bar{y} = (y / w_y) + 1$.
			(c) Same as (b) for $N_s = 10$.
		}
		\label{fig:kb}
	\end{center}
\end{figure}

Finally, we considered a four-QPC array.
To show the results more clearly, we
simulated the zero-temperature limit, where a single energy $E$ contributes.
As illustrated by Fig.~\ref{fig:kb}(a),
the interference pattern begins to resemble that of a diffraction grating. 
Near the QPCs, it exhibits the periodic refocusing known as the Talbot effect.
The repeat distance for the complete refocusing is~\cite{Rayleigh1881ocd}
\begin{equation}
y_T = \frac{\lambda}{1 - \sqrt{1 - \frac{\lambda^2}{a^2}}}\,,
\label{eqn:y_T}	
\end{equation}
where $\lambda = 2 \pi / k_y$ is the de Broglie wavelength of the IXs and $a$ is the distance between the QPCs.
For $\lambda = 100 \unit{nm}$ and $a = 1000\unit{nm}$,
this distance $y_T\simeq 20\,\mu\textrm{m}$
is beyond the range plotted in Fig.~\ref{fig:kb}(a).
Therefore, only the so-called fractional Talbot effect is
seen in that Figure.
Although too fine for conventional imaging,
these features may in principle be resolved by near-field optical techniques.
Note that for a grating with $N_s$ slits the crossover to the far-field diffraction occurs at the distance $\sim N_s y_T$,
which is prohibitively large for the transfer matrix simulations.
For a qualitative illustration of this crossover,
we computed the interference pattern simply by
adding a number of Gaussian beams, see Fig.~\ref{fig:kb}(b,c).

\textbf{Discussion and outlook}.
In this Letter, we considered a few prototypical examples of mesoscopic IX phenomena.
We analyzed the subband quantization of IX transport through a single QPC, the double-slit interference from two QPCs, and the Talbot effect in multiple QPCs.
As for electrons, these phenomena should be experimentally observable at low enough temperatures.
The present theory may be straightforwardly expanded to more complicated
potential landscapes and structures.

In the present work we neglected IX interactions.
This interaction is dipolar, which in 2D is classified as short-range,
parametrized by a certain interaction constant $g$.
As mentioned above,
the problem of the IX transport through a QPC is closely related to the problem 
of a BEC expansion from a harmonic trap.
Following the studies of the latter,~\cite{kagan1996evolution} interaction can be included at the mean-field level by using the Gross-Pitaevskii equation instead of the Schr\"odinger one.
For the single-QPC case, we expect the GPE solution to show a faster lateral expansion
of the exciton ``jet'', i.e., a more rapid growth of function $b(y)$.
For the double- and multi-QPC cases, we expect repulsive interaction to suppress
the interference fringes, similar to what is observed in experiments with cold atoms.~\cite{Kohstall2011}
The lineshape and amplitude (visibility) of the fringes remain to be investigated.

Applications of our theory to real semiconductor systems would also require taking into account 
spin and valley degrees of freedom of IXs.
It would be interesting to study 
spin transport of IXs~\cite{leonard2009spin, high2012spontaneous, finkelstein2017transition}
through nano-constrictions and associated spin textures.
This could be an alternative pathway to probing spin conductance of quasi-1D channels.~\cite{Meier2003mta}

The field of mesoscopic exciton systems is currently in its infancy but it is positioned
to grow, extending the phenomena studied in voltage-controllable electron systems to bosons. More importantly, it has the potential to reveal brand new phenomena.
There are many intriguing subjects for future work.

\medskip\noindent$\blacksquare$
AUTHOR INFORMATION

\noindent{}\textbf{Corresponding Author:}

\noindent{}$^*$E-mail: {chx035@ucsd.edu}

\noindent{}\textbf{Notes}

\noindent{}The authors declare no competing financial interest.

\noindent{}\textbf{Supporting Information Available:}

\noindent{}The Supporting Information includes:
i) electrostatic simulations of GaAs and TMD IX devices
ii) derivation of Eq.~\eqref{eqn:G}
iii) transport simulations by the transfer matrix method.
This material is available free of charge via the Internet at \url{http://pubs.acs.org}.


\medskip\noindent$\blacksquare$ ACKNOWLEDGEMENTS

\noindent{}The work at UCSD is supported by
the National Science Foundation under Grant ECCS-1640173 and by the
NERC, a subsidiary of Semiconductor Research Corporation
through the SRC-NRI Center for Excitonic Devices.

%
%




\providecommand{\latin}[1]{#1}
\makeatletter
\providecommand{\doi}
{\begingroup\let\do\@makeother\dospecials
	\catcode`\{=1 \catcode`\}=2 \doi@aux}
\providecommand{\doi@aux}[1]{\endgroup\texttt{#1}}
\makeatother
\providecommand*\mcitethebibliography{\thebibliography}
\csname @ifundefined\endcsname{endmcitethebibliography}
{\let\endmcitethebibliography\endthebibliography}{}

\newpage
\clearpage

\centerline{\Large{\textbf{Supporting information}}}

\setcounter{page}{1}
\renewcommand{\thepage}{{S}\arabic{page}}

\setcounter{equation}{0}
\renewcommand{\theequation}{{S}\arabic{equation}}

\setcounter{figure}{0}
\renewcommand{\figurename}{\textbf{Figure}}
\renewcommand{\thefigure}{{S}\arabic{figure}}

\renewcommand{\unit}[1]{\,\mathrm{#1}} 

\subsubsection{Electrostatic simulations}

To better assess the practicality of the proposed IX devices,
we numerically modeled several systems of the type shown in Fig.~1(a) of the main text.
The first example we considered is a TMD heterostructure
with the following layer sequence.
First, there is a thin crystal of hBN, which serves as
both the encapsulating layer and the gate dielectric.
Next, there are MoS$_2$ monolayers with two monolayers of hBN in between.
Finally, there is a thicker hBN capping layer.
Assuming the effective electron-hole separation $d$ is equal to the center-to-center distance of the two MoS$_2$ layers and using $0.33\unit{nm}$ and $0.65\unit{nm}$ for the inter-layer spacing of, respectively, hBN and MoS$_2$, we get $d = 1.31\unit{nm}$ in this design.
We took the thickness of the lower hBN layer to be $10\unit{nm}$, corresponding to about $30$ atomic layers, and the total thickness of the structure to be $d_g = 175\unit{nm}$.
The structure is assumed to reside on a ground plane, e.g.,
a graphite substrate, and to possess a set of top gates depicted in Fig.~1(c) of the main text.
Specifically, two symmetrically placed slits
shaped as square hairpins ($\sqSupset$ and $\sqSubset$),
divide the top plane into two rectangular side gates
maintained at voltage $V_g$ plus a narrow-necked central gate kept at voltage $V_d$.

To compute the electrostatic potential distribution in this system
we used a commercial finite-element solver.~\cite{Comsol}
Note that both hBN and MoS$_2$ are materials with anisotropic permittivities~\cite{Fogler2014}
$\epsilon_{x\mbox{-}y,\, \mathrm{hBN}} = 6.71$, $\epsilon_{z,\, \mathrm{hBN}} = 3.56$ and
$\epsilon_{x\mbox{-}y,\, \mathrm{MoS}_2} = 14.29$, $\epsilon_{z,\, \mathrm{MoS}_2} = 6.87$.
To simplify the simulations, we assigned MoS$_2$ the same permittivities as hBN.
In doing so, neglecting the difference of the in-plane permittivities is justified because the IX potential $U(x, y)$ is determined by the $E_z$-field.
However, it is necessary to account for the difference of the $z$-axis permittivities,
which we did by multiplying the calculated $U(x, y)$ by the factor
\begin{equation}
H + (1 - H) \frac{\epsilon_{z,\, \mathrm{hBN}}}{\epsilon_{z,\, \mathrm{MoS}_2}}\,
= 1.46\,.
\end{equation}
Here $H = {0.66\unit{nm}} / d = 0.50$
is the fraction of the distance $d$ occupied by the hBN spacer.
We found that a reasonably close fit to $U(x, y)$
plotted by the dashed line in Fig.~1(d) of the main text can be achieved 
if the width of the side gates is $100\unit{nm}$,
the width of the slits is $100\unit{nm}$ as well,
and the width of the neck of the central gate is $500\unit{nm}$.
Such dimensions are attainable with standard nanofabrication methods.
The applied voltages need to be $V_{d} = 4.1\unit{V}$ on the central gate and $V_g = 3.8\unit{V}$ on the side gates.

The second example we studied is an GaAs/AlGaAs heterostructure
composed of $100\unit{nm}$ of undoped AlGaAs,
followed by two $8$-$\mathrm{nm}$-thick GaAs quantum wells
separated by a $4$-$\mathrm{nm}$-thick AlGaAs barrier,
capped by another undoped AlGaAs layer to
the total thickness of $d_g = 500\unit{nm}$.
The center-to-center quantum well distance in this case is $d = 12\unit{nm}$.
The usual choice of the bottom electrode is an $n^+$-GaAs substrate.
We treated both GaAs and AlGaAs as isotropic materials with
identical permittivity of $\epsilon = 12.5$.
We found that the potential shown by the solid line in Fig.~1(d)
can be realized for the following electrode
dimensions: width $300\unit{nm}$, slit width
$200\unit{nm}$, central neck width $1100\unit{nm}$.
The applied voltages are $V_{d} = 2.00\unit{V}$ and $V_g = 1.85\unit{V}$.
Realizing this design again appears to be a straightforward task.

\begin{figure}
	\begin{center}
		\includegraphics[width=2.3in]{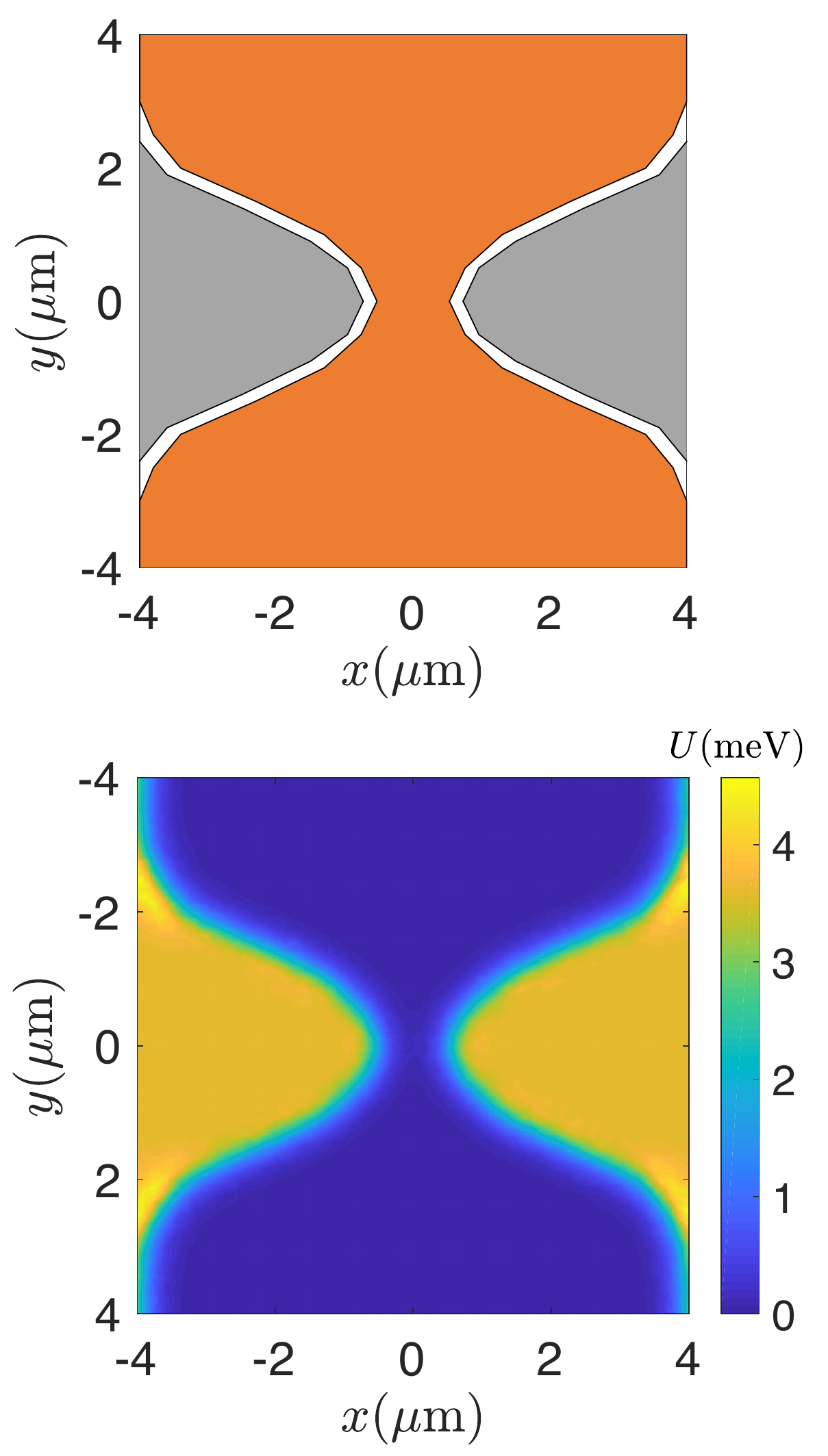}
		\caption{
			An example of simulated QPC devices.
			(a) top gate geometry
			(b) IX potential $U(x, y)$. In the proposed GaAs device (see text),
			this potential is realized for applied voltages
			$V_{d} = 2.00\unit{V}$ and $V_g = 1.85\unit{V}$.
		}
		\label{fig:gate}
	\end{center}
\end{figure}

Finally, we simulated the potential produced in the same
GaAs/AlGaAs heterostructure by a different top electrode pattern,
in which the slits delineating the side electrodes are curved,
Fig.~\ref{fig:gate}(a).
The computed $U(x, y)$ is plotted in Fig.~\ref{fig:gate}(b).
This design gives a better approximation to an ideal adiabatic QPC, see below.
\begin{figure}
	\begin{center}
		\includegraphics[width=2.0in]{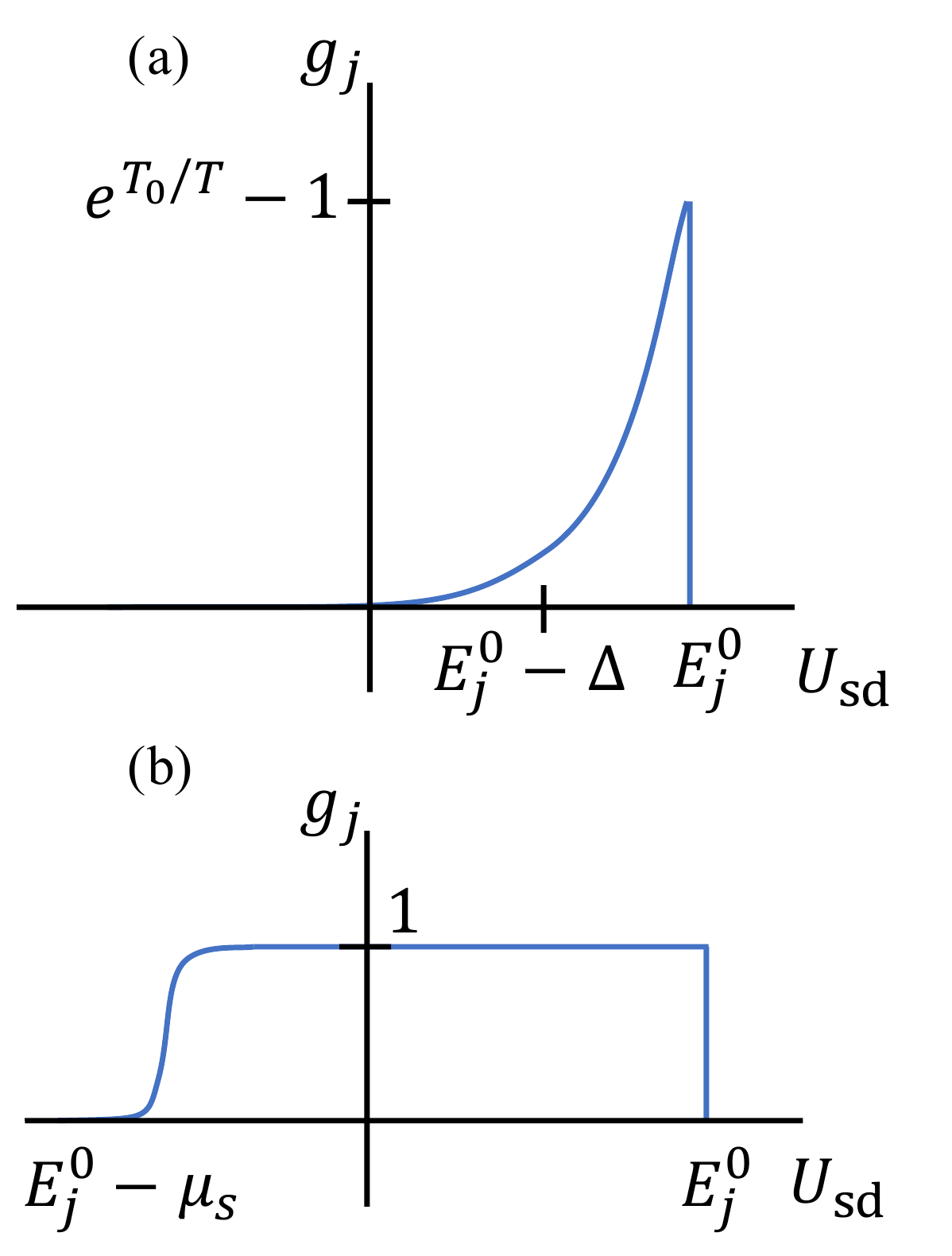}
		\caption{
			Dimensionless differential conductance of a single subband $j$ of a long (adiabatic) QPC:
			(a) bosonic case, $\Delta = \min (T, -\mu_s)$. (b) fermionic case.
		}
		\label{fig:g}
	\end{center}
\end{figure} 

\subsubsection{Derivation of Eq.~({\color{blue}12})}

The quasiparticle velocity of $j$th subband is
\begin{equation}
v_j(k_y) = \frac{1}{\hbar}\, \frac{d E_j}{d k_y}\,.
\end{equation}
Therefore, the current injected by the source into the subband $j$ is
\begin{align}
I_{s, j} &= {N}\int\limits_{0}^{\infty} t_j(E_j) f_{B, F}(E_j - \zeta_s)
\theta(E_j - U_{\rm{sd}}) v_j(k_y) \frac{d k_y}{2\pi}
\notag\\
&= \frac{N}{h}\int\limits_{U_{\rm{sd}}}^{\infty} t_j(E) f_{B, F}(E - \zeta_s) d E\,,
\label{eqn:I_js}
\end{align}
which agrees with Eq.~({\color{blue}11}) of the main text.
Here $B$, $F$ label the bosonic and fermionic cases, respectively.
The current $I_{d, j}$ injected by the drain is computed similarly.
For an adiabatic QPC with the transmission coefficients
$t_j(E) = \theta(E - E_j^0)$,
the net current $I_j = I_{s, j} - I_{d, j}$ of $j$th subband is
\begin{equation}
\begin{split}
I_j &= \frac{N}{h}\int\limits_{U_{\rm{sd}}}^{\infty} \theta(E-E_j^0)f_{F,B} (E  - U_{\rm{sd}} - \mu_s) d E\\
&- \frac{N}{h}\int\limits_0^{\infty} \theta(E - E_j^0)f_{F,B}(E  - \mu_d) d E\,.
\end{split}
\label{eqn:I_j}
\end{equation}
Accordingly, the differential conductance defined by Eq.~({\color{blue}10}) of the main text
can be written as 
\begin{equation}
G = \sum_j {\partial I_j} / {\partial U_{\rm{sd}}} = ({N} / {h}) \sum\limits_j g_j\,,
\end{equation}
where
\begin{equation}
g_j = \theta(E_j^0 - U_{\rm{sd}}) f_{F,B} (E_j^0 - U_{\rm{sd}} - \mu_s)
\end{equation}  
is the dimensionless conductance of subband $j$.
This quantity depends on $U_{\rm{sd}}$ the same way as the 
mirror-reflected quasiparticle occupation factor $f_{B, F}$ until $U_{\rm{sd}}$ reaches $E_j^0$,
at which point $g_j$ drops to zero, see Fig.~\ref{fig:g}.
As explained in the main text, the reason for this cutoff behavior is that
the current contributed by channel $j$ remains constant if $U_{\rm{sd}}$
increases past the subband bottom $E_j^0$.
Therefore, it does not affect the differential conductance.
However, if the QPC is not perfectly adiabatic, the transmission coefficient $t_j(E)$ has a gradual rather than the sharp onset at $E_j^0$, and so the cutoff of $g_j$ at $U_{\rm{sd}} = E_j^0$ is also gradual (see below).
Finally, summing over all the subbands,
we obtain the total differential conductance shown in Fig.~3 of the main text.

\subsubsection{Transport simulations}

To calculate the conductance of a bosonic QPC, we used the transfer matrix method.~\cite{Usuki1995nao}
In this method the 2D Schr\"odinger equation with energy $E$ and potential $U(x, y)$
in the simulation region $-L / 2 < x, y < L / 2$ is represented by a tight-binding model on a square lattice with a suitably small lattice constant $l$.
We take $L = 8\,\mu \mathrm{m}$ and $l = 20\unit{nm}$,
giving us a simulation grid with $N_x = 400$ points along the $x$-axis.
The boundary conditions for the Schr\"odinger equation
correspond to semi-infinite leads
inside which the particles are subject to $y$-independent potentials
\begin{equation}
U_s(x) = U(x, -L / 2)\,,
\quad
U_d(x) = U(x, +L / 2)\,,
\label{eqn:U_s}
\end{equation}
for the source and the drain, respectively. 
The energy spectrum of the source lead consists of $N_x$ subbands with dispersion
$E_i^s + E_k$.
Here $E_i^s$ are the quantized energies of motion in the potential $U_s(x)$,
\begin{equation}
E_k =  \frac{\hbar^2}{m l^2} (1 - \cos k l),
\end{equation}
is the kinetic energy of the $y$-motion,
and $k$ is the $y$-axis momentum.
The $y$-axis velocity is
\begin{equation}
u_i(E_k) = \frac{1}{\hbar}\, \frac{d E_k}{d k}
= \sqrt{\frac{2 E_k}{m} \left(1 - \frac{m l^2}{2 \hbar^2}E_k \right)}\,.
\end{equation}
Similar energy spectrum exists in the drain.
The transfer matrix algorithm computes the $N_x \times N_x$ matrix of the transmission coefficients $t_{i j}(E)$
between every subband $i$ of the source and every subband $j$ of the drain.
To reduce the computation load, we did not explicitly include
the potential drop $U_{\mathrm{sd}}$ in $U(x, y)$.
As explained in the main text, this drop simply changes the particle velocity.
We accounted for it by including the ratio of velocities before and after
the drop as a multiplicative factor.
Accordingly,
our formula for the total current injected into the QPC by the subband $i$ of the source is
\begin{equation}
\begin{split}
I_i =& \sum\limits_{k} \frac{1}{L}
\frac{u_i(E_k)}{u_i(E_k + U_{\mathrm{sd}})}
f_B(E_k + E_i^s - \mu_s )\\
\times& \sum\limits_j t_{ij}(E_k + E_i^s + U_{\mathrm{sd}})
u_j(E_k + U_{\mathrm{sd}})\,.
\end{split}
\label{eqn:I_in}
\end{equation}
The $k$-grid in the summation must be dense enough to achieve accuracy.
We found the following choice adequate:
\begin{equation}
k = \frac{2\pi n}{L}, \quad n = 0, 1, \ldots, n_{\mathrm{max}} = \frac{L}{4 l}\,.
\end{equation}
The net current through the QPC for the case of empty drain is simply $\sum_i I_i$.

For consistency check, consider first an ideal adiabatic QPC, where the transmission coefficients are $t_{ij}(E) = \delta_{ij}\theta(E - E_j^0)$.
The rule for changing variables from $k$ to $k_y$ is found from the energy conservation
\begin{equation}
E_j(k_y) = E_k + E_i^s + U_{\mathrm{sd}}\,,
\end{equation}
which entails
\begin{equation}
\frac{1}{L} \sum\limits_{k}\ldots \to \int \frac{d k_y}{2\pi}\, \frac{v_j(E)}{u_i(E_k)} \ldots
\end{equation}
Using these expressions,
it is easy to see that Eq.~\eqref{eqn:I_in} is indeed transformed to Eq.~\eqref{eqn:I_js} in this limit.
The corresponding analytical results for the current and conductance as functions of $U_{\mathrm{sd}}$ are plotted in Fig.~\ref{fig:g} by the thick lines.
Shown in the same Figure by the dashed lines are our numerical results for the potential
of Fig.~\eqref{fig:gate}(b).
They prove to be fairly close to the adiabatic limit except for some rounding of the conductance peaks and the current steps, as expected.
On the other hand,
the results for the potential $U(x, y)$ given by Eq.~({\color{blue}1}) of the main text,
which are indicated by the thin lines,
deviate more from the ideal case.

\begin{figure}
	\begin{center}
		\includegraphics[width=3.3in]{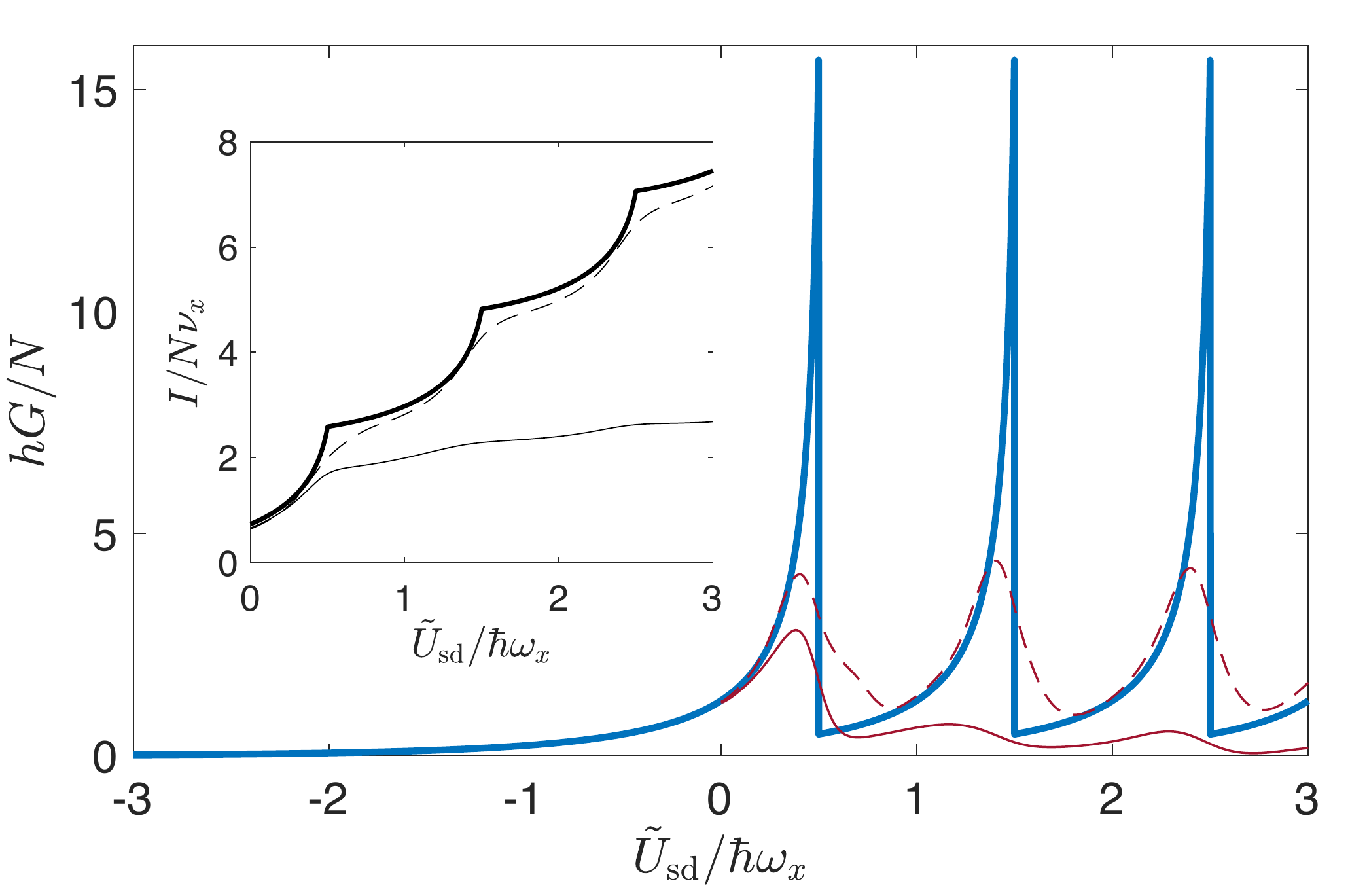}
		\caption{
			Conductance of a bosonic QPC
			as a function of electrostatic potential energy difference between the source and the QPC center $\tilde{U}_{\mathrm{sd}} \equiv U_{\mathrm{sd}} - U(0, y_0)$.
			The three lines correspond to three models considered:
			an ideal adiabatic constriction (thick line),
			a QPC with the potential of Fig.~\ref{fig:gate} (dashed line),
			a QPC defined by Eq.~({\color{blue}1}) of the main text (thin line).
			Inset: the total current \latin{vs}. $\tilde{U}_{\mathrm{sd}}$ in these three models.
			Parameters: $T = 0.8 \hbar\omega_x$, $\mu_s = -0.05 \hbar\omega_x$.
		}
		\label{fig:g}
	\end{center}
\end{figure}

\providecommand{\latin}[1]{#1}
\makeatletter
\providecommand{\doi}
{\begingroup\let\do\@makeother\dospecials
	\catcode`\{=1 \catcode`\}=2 \doi@aux}
\providecommand{\doi@aux}[1]{\endgroup\texttt{#1}}
\makeatother
\providecommand*\mcitethebibliography{\thebibliography}
\csname @ifundefined\endcsname{endmcitethebibliography}
{\let\endmcitethebibliography\endthebibliography}{}

\end{document}